\documentclass[showpacs,preprintnumbers,amsmath,amssymb,showkeys]{revtex4}

\usepackage{graphicx}
\usepackage{dcolumn}
\usepackage{bm}

\usepackage{multirow}

\begin{document}

\title{$\overline pD$ atoms in models of realistic potentials}
\author{Y. Yan}
\email{yupeng@sut.ac.th}
\author{K. Khosonthongkee}
\author{C. Kobdaj}
\author{P. Suebka}
\affiliation{School of Physics, Suranaree University of
Technology, 111 University Avenue, Nakhon Ratchasima 30000,
Thailand}

\date{\today}

\begin{abstract}
The $\overline pD$ atoms are studied in models of various realistic,
popular $\overline NN$ potentials. The small energy shifts and decay
widths of the atoms, which stem from the short-ranged strong
interactions between the antiproton and deuteron, are evaluated in a
well-established, accurate approach based on the Sturmian functions.
The investigation reveals that none of the employed potentials,
which reproduce the $\overline NN$ scattering data quite well, is
able to reproduce the experimental data of the energy shifts of the
$2p$ $\overline pD$ atomic states. The energy shifts of the $2p$
$\overline pD$ atomic states are very sensitive to the $\overline
NN$ strong interactions, hence the investigation of the $\overline
pD$ atoms is expected to provide a good platform for refining the
$\overline NN$ interaction, especially at zero energy.
\end{abstract}


\maketitle

\section{Introduction}
The second simplest antiprotonic atom is the antiprotonic deuteron
atom $\overline pD$, consisting of an antiproton and a deuteron
bound mainly by the Coulomb interaction but distorted by the short
range strong interaction. The study of the $\overline pD$ atom is
much later and less successful than for other exotic atoms like
the protonium and pionium. Experiments were carried out at LEAR
just in very recent years to study the properties of the
$\overline pD$ atom \cite{expt1,expt2}. Even prior to the experiments
some theoretical works \cite{theo1,theo2,theo3} had been carried out to study
the $\overline pD$ atomic states in simplified $\overline pD$
interactions. Recently, a theoretical work \cite{theo4} proposed
a mechanism explaining the unexpected behavior, of the scattering lengths of
$\overline NN$ and $\overline pD$ system, that
the imaginary part of the scattering length does not increase with
the size of the nucleus.

In the theoretical sector, one needs to overcome at least two
difficulties in the study of the $\overline pD$ atom.
First, the interaction between the antiproton and the deuteron
core should be derived from realistic $\overline NN$ interactions,
for example, the Paris $\overline NN$ potentials
\cite{Paris94,Paris99,Paris04}, the Dover-Richard $\overline NN$
potentials I (DR1) and II (DR2) \cite{DR1,DR2},
and the Kohno-Weise $\overline NN$ potential \cite{KW}. Even if a reliable $\overline pD$
interaction is in hands, the accurate evaluation of the energy
shifts and decay widths (stemming for the strong $\overline pD$
interactions) and especially of the nuclear force distorted wave
function of the atom is still a challenge. It should be pointed
out that the methods employed in the works
\cite{theo1,theo2,theo3} are not accurate enough for evaluating
the wave functions of the $\overline pD$ atoms.

In the present work we study the $\overline pD$ atom problem
employing a properly adapted numerical method based on Sturmian
functions \cite{stur3}. The method accounts for both the strong
$short$ range nuclear potential (local and non-local) and the $long$
range Coulomb force and provides directly the wave function of the
$\overline pD$ system with complex eigenvalues
$E=E_R-i\frac{\Gamma}{2}$. The protonium and pionium problems have
been successfully investigated \cite{yanatom,yanpionium} in the
numerical approach. The numerical method is much more powerful,
accurate and much easier to use than all other methods applied to
the exotic atom problem in history. The $\overline pD$ interactions
in the work are derived from various realistic $\overline NN$
potential, which is state-dependent. The work is organized as
follows. The $\overline pD$ interactions are expressed in Sec. II in
terms of the $\overline NN$ interactions. In Sec. III the energy
shifts and decay widths of the $1s$ and $2p$ $\overline pD$ atomic
states are evaluated. Discussions and conclusions are given in Sec.
III, too.

\section{$\overline pD$ interactions in terms of $\overline NN$ potentials}
We start from the Schr\"odinger equation of the
antiproton-deuteron system in coordinate space
\begin{equation}\label{eq::1}
\left(\frac{P^2_\rho}{2M_\rho}+\frac{P^2_\lambda}
{2M_\lambda}+V_{12}(\vec r_2- \vec r_1)+V_{13}(\vec r_3- \vec
r_1)+V_{23}(\vec r_3- \vec r_2)\right)
\Psi(\vec\lambda,\vec\rho)=E\Psi(\vec\lambda,\vec\rho)
\end{equation}
where $\vec\lambda$ and $\vec\rho$ are the Jacobi coordinates of
the system, defined as
\begin{equation}\label{eq::2}
\vec\lambda=\vec r_3-\frac{\vec r_1+\vec r_2}{2},\;\;\;
\vec\rho=\vec r_2-\vec r_1
\end{equation}
$M_\rho=M/2$ and $M_\lambda=2M/3$ are the reduced masses. Here we
have assigned, for simplicity, the proton and neutron the same mass
$M$. Eq. (\ref{eq::1}) can be expressed in the form, where the
strong interaction is expressed in the isospin basis,
\begin{equation}\label{eq::3}
\left(\frac{P^2_\rho}{2M_\rho}+\frac{P^2_\lambda}
{2M_\lambda}+V_{S}+V_{C}\right)
\Psi(\vec\lambda,\vec\rho)=E\Psi(\vec\lambda,\vec\rho)
\end{equation}
where $V_S$ and $V_C$ stand for the nuclear interaction and
Coulomb force, respectively, and take the forms
\begin{equation}\label{eq::4}
V_S=V^0_{NN}(\vec r_2-\vec r_1)+\frac{1}{4}[V^0_{\overline
NN}(\vec r_3-\vec r_1)+V^0_{\overline NN}(\vec r_3-\vec r_2)] +
\frac{3}{4}[V^1_{\overline NN}(\vec r_3-\vec r_1)+V^1_{\overline
NN}(\vec r_3-\vec r_2)]
\end{equation}
\begin{equation}\label{eq::5}
V_C=\frac{1}{2}[V_C(\vec r_3-\vec r_1)+V_C(\vec r_3-\vec r_2)]
\end{equation}
$V^0$ and $V^1$ in eqs. (\ref{eq::4}) and (\ref{eq::4}) are the isospin 0 and 1 nuclear
interactions, respectively. Note that we have assigned $\vec r_{12}$ as the relative
coordinate of the deuteron core.

One may express the interactions $V_C$ and $V_S$ in eq. (\ref{eq::4})
eq. (\ref{eq::5}) in terms of the interactions of certain $\overline NN$ states.
In the $\left|JMLS\right\rangle$ basis of the $\overline pD$
states
\begin{equation}\label{eq::6}
\left|JMLS\right\rangle =\left|(L_\rho\otimes L_\lambda)_L\otimes
 (S_{12}\otimes S_3)_S\right\rangle_{JM}
\end{equation}
we derive
\begin{equation}\label{eq::7}
\left(H_0+W_C(\lambda,\rho)+V^0_{NN}(\rho)+W_S(\lambda,\rho)\right)
\Psi(\lambda,\rho)=E\Psi(\lambda,\rho)
\end{equation}
with
\begin{eqnarray}\label{eq::8}
H_0=\frac{P^2_\rho}{2M_\rho}+\frac{P^2_\lambda}
{2M_\lambda}
\end{eqnarray}
$W_C$ and $W_S$ in eq. (\ref{eq::7}) are respectively the Coulomb force and
strong interaction between the antiproton and deuteron, and
$V^0_{NN}$ the interaction between the proton and neutron in the
deuteron core. $W_C$ and $W_S$ are derived explicitly as
\begin{eqnarray}\label{eq::9}
W_C(\lambda,\rho)=\frac{1}{2}\int^1_{-1}dx\,V_C(r_{13})
\end{eqnarray}
\begin{eqnarray}\label{eq::10}
W_S(\lambda,\rho)=\frac{1}{2}\int^{1}_{-1}dx\,\sum_{Q,Q'}
\left\langle P|Q\right\rangle\left\langle Q
\right|\,V_{\overline NN}(\vec r_{13})\,\left|Q'\right\rangle\left\langle Q'|P'\right\rangle
\end{eqnarray}
with \begin{eqnarray}\label{eq::11}
V_{\overline NN}(\vec r_{13})= \frac{1}{2}V^0_{\overline NN}(\vec
r_{13})+\frac{3}{2}V^1_{\overline NN}(\vec r_{13})
\end{eqnarray}
\begin{eqnarray}\label{eq::12}
r_{13}\equiv |\vec r_1-\vec
r_3|=\left(\lambda^2+\rho^2/4-\lambda\rho\,x\right)^{1/2}
\end{eqnarray}
where $x=\cos\theta$ with $\theta$ being the angle between $\vec\lambda$ and $\vec\rho$.
In eq. (\ref{eq::10}) $|P\rangle\equiv\left|JMLS\right\rangle$ and
$|P'\rangle\equiv\left|JML'S\right\rangle$ are as defined in eq. (\ref{eq::6})
while the states $\left|\,Q\right\rangle$ and $\left|\,Q'\right\rangle$ are
\begin{equation}\label{eq::13}
\left|\,Q\right\rangle =\left|(L_\sigma\otimes S_{13})_{J_\sigma}\otimes
 (L_\gamma\otimes S_2)_{J_\gamma}\right\rangle_{JM}
\end{equation}
\begin{equation}\label{eq::14}
\left|\,Q'\right\rangle =\left|(L'_\sigma\otimes S_{13})_{J_\sigma}\otimes
 (L_\gamma\otimes S_2)_{J_\gamma}\right\rangle_{JM}
\end{equation}
Here $\vec\sigma$ and $\vec\gamma$ are also the Jacobi coordinates of
the system, defined as
\begin{equation}\label{eq::15}
\vec\gamma=\vec r_2-\frac{\vec r_1+\vec r_3}{2},\;\;\;
\vec\sigma=\vec r_3-\vec r_1
\end{equation}
So defined the states $\left|\,Q\right\rangle$ and
$\left|\,Q'\right\rangle$ is based on the consideration that the
$\overline NN$ interactions can be easily expressed in the
$\left|J_\sigma M_\sigma L_\sigma S_{13}\right\rangle$ basis of
the $\overline NN$ states. Note that $\langle P|Q\rangle$ depends
on not only the quantum numbers of the states $|P\rangle$ and
$|Q\rangle$, but also $\lambda$, $\rho$ and the angle $\theta$
between $\vec\lambda$ and $\vec\rho$ resulting from the projection
of the orbital angular momenta between different Jacobi
coordinates. We listed the integral kernels in eq.
(\ref{eq::10}), $\sum_{Q,Q'} \left\langle
P|Q\right\rangle\left\langle Q \right|\,V(\vec
r_{13})\,\left|Q'\right\rangle\left\langle Q'|P'\right\rangle$,
for the lowest $\overline pD$ states in the approximation that the
deuteron core is assumed in the S-state, as follows:
\begin{eqnarray}\label{eq::16}
\begin{array}{ll}
|P\rangle=|P'\rangle=|^{2}S_{1/2}\rangle:\;\;\;& \frac{3}{4}V_{\overline NN}(^1S_0)+\frac{1}{4} V_{\overline NN}(^3S_1)
\\
|P\rangle=|P'\rangle=| ^{4}S_{3/2}\rangle:\;\;\;& V_{\overline NN}(^3S_1)
\\
|P\rangle=|P'\rangle=|^{2}P_{1/2}\rangle:\;\;\;& F_1^2\cdot\left[\frac{1}{12}V_{\overline
NN}(^3P_0)+\frac{3}{4}V_{\overline NN}(^1P_1)
 +\frac{1}{6}V_{\overline NN}(^3P_1)\right]
\\
|P\rangle=|P'\rangle=| ^{4}P_{1/2}\rangle:\;\;\;& F_1^2\cdot\left[\frac{2}{3}V_{\overline NN}(^3P_0)
 +\frac{1}{3}V_{\overline NN}(^3P_1)\right]
\\
|P\rangle=|P'\rangle=| ^{2}P_{3/2}\rangle:\;\;\;& F_1^2\cdot\left[\frac{3}{4}V_{\overline NN}(^1P_1)
 +\frac{1}{24}V_{\overline NN}(^3P_1)
 +\frac{5}{24}V_{\overline NN}(^3P_2)\right]
\\
|P\rangle=|P'\rangle=| ^{4}P_{3/2}\rangle:\;\;\;& F_1^2\cdot\left[\frac{5}{6}V_{\overline NN}(^3P_1)
 +\frac{1}{6}V_{\overline NN}(^3P_2)\right]
\\
|P\rangle=|P'\rangle=| ^{4}P_{5/2}\rangle:\;\;\;& F_1^2\cdot V_{\overline NN}(^3P_2)
\\
|P\rangle=|P'\rangle=|  ^{4}D_{3/2}\rangle:\;\;\;& F_3^2\cdot\left[\frac{1}{2}V_{\overline NN}(^3D_1)
  +\frac{1}{2}V_{\overline NN}(^3D_2)\right]
\\
|P\rangle=|P'\rangle=| ^{2}F_{3/2}\rangle:\;\;\;& F_2^2\cdot V_{\overline NN}(^3F_2)
\\
|P\rangle=|P'\rangle=|    ^{4}F_{5/2}\rangle:\;\;\;& F_2^2\cdot\left[\frac{4}{9}V_{\overline NN}(^3F_2)
+\frac{5}{9}V_{\overline NN}(^3F_3)\right]
\\
|P\rangle=| ^{4}P_{3/2}\rangle,\; |P'\rangle=|^{4}F_{3/2}\rangle:\;\;\;& F_1F_2\cdot \frac{1}{\sqrt{6}}V_{\overline NN}(^3PF_2)
\\
|P\rangle=| ^{4}P_{5/2}\rangle,\; |P'\rangle=|^{4}F_{5/2}\rangle:\;\;\;& F_1F_2\cdot \frac{2}{3}V_{\overline NN}(^3PF_2)
\\
|P\rangle=| ^{4}S_{3/2}\rangle,\; |P'\rangle=|^{4}D_{3/2}\rangle:\;\;\;
 & F_3\cdot\left[\frac{1}{\sqrt{2}}V_{\overline NN}(^3SD_1)
  +\frac{1}{\sqrt{2}}V_{\overline NN}(^3SD_2)\right]
\end{array}
\end{eqnarray}
where $|P\rangle\equiv\left|JMLS\right\rangle$ and
$|P'\rangle\equiv\left|JML'S\right\rangle$ are the $\overline pD$ atomic states.
Both the $\overline pD$ and $\overline NN$ states in eq. (\ref{eq::16})
are labelled as $^{2S+1}L_J$ with $S$, $L$ and $J$ being respectively
the total spin, total orbital angular momentum and total angular
momentum. The potentials $V_{\overline NN}$,
being functions of $r_{13}
= \sqrt{\lambda^2+\rho^2/4-\rho\lambda x}$, stand for the
$\overline NN$ interactions for various $\overline NN$ states as
indicated in the brackets.

The $F_1$, $F_2$ and $F_3$ in eq. (\ref{eq::16}) are
functions of only $\lambda$ and $\rho$, taking the forms
\begin{eqnarray}\label{eq::17}
F_1=\left\{
\begin{array}{lll}
1-\frac{1}{12}\frac{\rho^2}{\lambda^2},
&\rho<2\lambda \\
\frac{4\lambda}{3\rho},&\rho>2\lambda \\
\end{array}
\right.
\end{eqnarray}
\begin{eqnarray}
F_2=\left\{
\begin{array}{lll}
\left(1-\frac{\rho^2}{4\lambda^2}\right)^2,
&\rho<2\lambda \\
0,&\rho>2\lambda \\
\end{array}
\right.
\end{eqnarray}
\begin{eqnarray}\label{eq::17}
F_3=\left\{
\begin{array}{lll}
_2F^1(1,-\frac{3}{2},\frac{3}{2},\frac{\rho^2}{4\lambda^2}),
&\rho<2\lambda \\
\frac{5}{8}-\frac{3\rho^2}{32\lambda^2}+
\mbox{Artanh}\left(\frac{2\lambda}{\rho}\right)
\left[\frac{3\lambda}{4\rho}-\frac{3\rho}{8\lambda}+\frac{3\rho^3}{64\lambda^3}\right],&\rho>2\lambda \\
\end{array}
\right.
\end{eqnarray}
where $_2F^1(\alpha,\beta,\gamma,x)$ is the hypergeometric
function and $\mbox{Artanh}(x)$ the inverses hyperbolic tangent function.
\section{Energy shifts and decay widths of $\overline pD$ atoms}
It is not a simple problem to accurately evaluate the energy
shifts and decay widths, especially wave functions of exotic atoms
like protonium, pionium and antiproton-deuteron atoms, which are
mainly bound by the Coulomb force, but also effected by the
short range strong interaction. In this work we study the
$\overline pD$ atoms in the Sturmian function approach which has been
successfully applied to our previous works \cite{yanatom,yanpionium}. Employed for the
$\overline NN$ interactions are various realistic $\overline NN$
potentials, namely, the Paris $\overline NN$ potentials of the 1994 version (Paris84),
1998 version (Paris98) and
2004 version (Paris04), the Dover-Richard $\overline NN$ potentials I (DR1) and II (DR2),
and  the Kohno-Weise
$\overline NN$ potential (KW). In this preliminary work, we just limit our study to
the approximation of undistorted deuteron core. However, one may see that
the main conclusions of the work are free of this approximation.

Shown in Table I are the energy shifts and decay widths, which stem
from the Paris98, DR2 and KW $\overline NN$ interactions, in the approximation of undistorted deuteron
core. The theoretical results for other interactions like Paris84, Paris04 and DR1 are quite
similar to the ones listed in Table I. The wave function of the undistorted deuteron
core is evaluated in the Bonn OBEPQ potential \cite{Machleidt}.
It is found that the theoretical results for the $1s$ $\overline pD$ atomic states are
more or less the same by all the employed $\overline NN$ potentials. The predicted energy
shifts are roughly as twice large as the experimental data. However, one may expect
that the predictions of the potentials in question could be improved to some extent by solving the
$\overline pD$ dynamical equation in eq. (\ref{eq::7}) without any approximation.
A better treatment of the deuteron core will yield lower $1s$ $\overline pD$ atomic states,
hence smaller energy shifts. The theoretical results
for the decay widths of the $1s$ $\overline pD$
atoms are also larger than the experimental data though not as far from the data as for
the energy shifts.
The predictions for the decay widths are also expected to be improved by treating the deuteron
core more properly.
\begin{table}
\begin{center}
\begin{tabular}{|c|c|c|c|c|c|c|c|c|}
\hline
& \multicolumn{2}{c|}{Paris98} & \multicolumn{2}{c|}{DR2} & \multicolumn{2}{c|}{KW} & \multicolumn{2}{c|}{Data}\\
 \cline{2-9}
 &$\Delta E$& $\Gamma$ & $\Delta E$& $\Gamma$ & $\Delta E$& $\Gamma$& $\Delta E$& $\Gamma$\\
 \hline
$^{2}S_{1/2}$ & -2445  &1781     & -2673&2380    &-2478&2450   & &  \\
\hline
 $^{4}SD_{3/2}$ & -2680&2822     & -2668&2390    &-2503&2469   & & \\
 \hline
 $^{2}P_{1/2}$ & -186  &584      & 17   &896     &99   &657    & & \\
 \hline
 $^{4}P_{1/2}$ & 265   &402      & 47   &846     &101  &785    & & \\
 \hline
 $^{2}P_{3/2}$ & -128  &515      & 14   &897     &98   &643    & &\\
 \hline
 $^{4}PF_{3/2}$ & 282  &477      & 21   &887     &97   &648    & & \\
 \hline
 $^{4}PF_{5/2}$ & 244  &814      & 21   &877     &101  &660    & &\\
\hline
\multirow{2}{*}{$\overline {\Delta E}_{1s}$, $\overline {\Gamma}_{1s}$} & -2602&2475 & -2670 &2387&-2494 &2463 &
$-1050\pm 250$\cite{expt1}& $1100\pm 750$\cite{expt1}\\
& & & & & & & & $2270 \pm 260$\cite{expt2}  \\
\hline
$\overline {\Delta E}_{2p}$, $\overline {\Gamma}_{2p}$ & 124&602 &22 & 883 &99 &668&
$-243\pm 26$\cite{expt2}& $489\pm 30$\cite{expt2} \\
 \hline
\end{tabular}
\caption{The energy shifts $\Delta E$ and decay widths of the $1s$ and $2p$ antiproton-deuteron
atomic states in the approximation of undistorted deuteron core. The minus sign of the energy shifts means that the
strong interaction is repulsive. The units are eV and meV for $1s$ and $2p$ states, respectively.}
\end{center}
\end{table}

The theoretical predictions for the energy shifts of the $2p$ $\overline pD$ atomic states
are totally out of line for all the $\overline NN$ potentials employed. The experimental data show
that the averaged energy level of the $2p$ $\overline pD$ atoms is pushed up by the strong
interaction, the same as for the $1s$ $\overline pD$ atoms, but the theoretical results
uniquely show the averaged energy level shifting down. It is unlikely to improve, by treating
the deuteron core more accurately, the theoretical
predictions of the $\overline NN$ potentials in question for the $2p$ $\overline pD$ energy shifts
since a more accurate treatment of the deuteron core will lead to deeper $2p$ $\overline pD$ atomic states.

All the $\overline NN$ potentials employed in the work reproduce $\overline NN$ scattering data reasonably, but
badly fail to reproduce the energy shifts of the $2p$ $\overline pD$ atoms.
The investigation of the $\overline pD$ atoms
may provide a good platform for refining the $\overline NN$ interaction, especially at zero energy since
the energy shifts of the $2p$ $\overline pD$ atomic states are very sensitive to
the $\overline NN$ strong interactions.

The research here is just a preliminary work, where a frozen, S-state deuteron is employed.
The work may be improved at two steps, considering that the numerical evaluation is time-consuming.
One may, at the first step, solve the $\overline pD$ dynamical equation in eq. (\ref{eq::7}) by expanding
the $\overline pD$ wave function in a bi-wave basis of the Sturmian functions, where a realistic nucleon-nucleon
potential is employed but the deuteron core is assumed to be at the S-state. Such an evaluation is still
manageable at a personnel computer but it may take a week or longer. We may compare the results of the improved
work with the results here to figure out how important an unfrozen deuteron core is.

One may also consider, at the second step, to solve the $\overline pD$ dynamical equation in eq. (\ref{eq::7})
by expanding the $\overline pD$ wave function in a bi-wave basis of the Sturmian functions without any approximation,
where realistic nucleon-nucleon and nucleon-antinucleon potentials are employed and the deuteron core is allowed to
be at both the S- and D-waves. It is certain that the numerical calculation will take longer time
but, anyway, we will do it after we complete the first-step improvement.

\section*{Acknowledgements}

We thank Th. Gutsche, Amand Faessler and V.E. Lyubovitskij,
Institute for Theoretical Physics, T\"ubingen University for helpful
discussions. The work is supported in part by the National Research
Council of Thailand through Suranaree University of Technology and
the Commission on Higher Education, Thailand (CHE-RES-RG Theoretical
Physics).

\end{document}